\documentstyle[pra,epsf,aps,twocolumn]{revtex}
\begin{document}
\title{Torsion Dilaton and Novel Minimal Coupling Principle}
\author{P.~P.~Fiziev \thanks{ E-mail:\,\, fiziev@phys.uni-sofia.bg}}
\address{Department of Theoretical Physics, Faculty of Physics,
Sofia University, 5 James Bourchier Boulevard, Sofia~1164, Bulgaria
\thanks{Permanent address}\\
and\\
Bogoliubov Laboratory of Theoretical Physics,
Joint Institute for Nuclear Research, 141980 Dubna, Moscow Region, Russia}
\maketitle
\begin{abstract}
We propose a novel self consistent minimal coupling principle in presence of
torsion dilaton field. This principle yields a new local dilatation symmetry
and predicts the interactions of torsion dilaton with the real matter and with
metric. The soft violation of this symmetry yields a physical dilaton and
a simple relation between Cartan scalar curvature and cosmological constant
in this new model of gravity with propagating torsion. Its relation with
scalar-tensor theories of gravity and a possible use of torsion dilaton
in the inflation scenario is discussed.

\noindent{PACS number(s): 04.50.+h, 04.40.Nr, 04.62.+v}
\end{abstract}
\sloppy
\renewcommand{\baselinestretch}{1.3} %
\newcommand{\sla}[1]{{\hspace{1pt}/\!\!\!\hspace{-.5pt}#1\,\,\,}\!\!}
\newcommand{\db}{\,\,{\bar {}\!\!d}\!\,\hspace{0.5pt}}
\newcommand{\lambdab}{\,\,{\bar {}\!\!\lambda}\!\,\hspace{0.5pt}}
\newcommand{\partb}{\,\,{\bar {}\!\!\!\partial}\!\,\hspace{0.5pt}}
\newcommand{\dsla}{\partb}
\newcommand{\Boxb}{\Box^{\hskip -7.7pt {}^{-}}}
\newcommand{\BoxD}{\Box^{\hskip -7.2pt {}^{{}^{D}}}}
\newcommand{\nablaD}{\nabla^{\hskip -7.2pt {}^{{}^{D}}}}
\newcommand{\GammaD}{\Gamma^{\hskip -6.2pt {}^{{}^{D}}}}
\newcommand{\SigmaD}{\Sigma^{\hskip -7.pt {}^{{}^{D}}}}
\newcommand{\ThetaD}{\Theta^{\hskip -7.pt {}^{{}^{D}}}}
\newcommand{\LambdaD}{\Lambda^{\hskip -7.pt {}^{{}^{D}}}}
\newcommand{\AD}{{\stackrel{\hskip 5.pt{}_{\it D}}{A}}}
\newcommand{\DD}{{\stackrel{\hskip 3.5pt{}_{\it D}}{D}}}
\newcommand{\ND}{{\stackrel{\hskip 2.7pt{}_{\it D}}{N}}}
\newcommand{\RD}{{\stackrel{\hskip 2.7pt{}_{\it D}}{R}}}
\newcommand{\SD}{{\stackrel{\hskip 3.7pt{}_{\it D}}{S}}}
\newcommand{\lfrac}[2]{{#1}/{#2}}
\newcommand{\sfrac}[2]{{\small \,\,\hbox{${\frac {#1} {#2}}$}}}
\newcommand{\ben}{\begin{eqnarray}}
\newcommand{\een}{\end{eqnarray}}
\newcommand{\la}{\label}
%
\section{Introduction.}
Recently it was recognized \cite{Saa}-\cite{F2} a necessity of a new minimal
coupling principle (MCP) in affine-metric four-dimensional space-times
${\cal M}^{4}\{g_{\alpha\beta}(x), \Gamma_{\alpha\beta}{}^\gamma(x)\}$\,\,
with curvature ${R_{\alpha\beta\mu}}^\nu := 2\left(
\partial_{[\alpha} \Gamma_{\beta]\mu}{}^\nu +
\Gamma_{[\alpha|\sigma|}{}^\nu\Gamma_{\beta]\mu}{}^\sigma\right)\neq 0$,
torsion $S_{\alpha\beta}{}^\gamma := \Gamma_{[\alpha\beta]}{}^\gamma\neq 0$
and nonmetricity
$N_{\alpha\beta\gamma}:=\nabla_\alpha g_{\beta\gamma} \equiv 0$,
equipped with pseudo-Riemannian $(+,-,-,-)$ metric:
$g_{\alpha\beta} = g_{\beta\alpha}$,
$g=\det ||g_{\alpha\beta}|| \neq 0$,
$g^{\alpha\gamma}g_{\gamma\beta}=\delta^\alpha_\beta$,
interval $ds^2=g_{\alpha\beta} dx^\alpha dx^\beta$,
and with affine-metric connection with coefficients
$\Gamma_{\alpha\beta}{}^\gamma = {\gamma \brace \alpha \beta}
+{S_{\alpha\beta} }^\gamma+{S^\gamma}_{\alpha\beta}+{S^\gamma}_{\beta\alpha}$,
${\gamma \brace \alpha \beta}= {\sfrac 1 2}g^{\gamma \mu}
(\partial_\alpha g_{\mu\beta} +\partial_\beta g_{\mu\alpha} -
\partial_\mu g_{\alpha\beta})$
being the Christoffel symbols \cite{SKN}.
The symmetric part
$\Gamma_{ \{\alpha\beta\}}{}^\gamma =
{\gamma \brace \alpha \beta}+{S^\gamma}_{\alpha\beta}+{S^\gamma}_{\beta\alpha}$
depends on the torsion because of the metricity condition.

A new MCP is needed  to overcome the so called G-A problem:
in presence of torsion the standard MCP
yields different results when applied in action principle,
or directly in the equations of motion for particles and fields
(see for example \cite{Saa} -\cite{F2} and the references therein).
This turns us back to the old question what are the right equations of
motion for a free test particles and fields in presence of torsion
\cite{PH}.
For example,
a free test particle may move according to Newton's law of inertia
(MCP applied directly in equations of motion)
with zero absolute acceleration along autoparallel (A) lines
\ben
m\,a^\gamma =
mc^2\left( {\frac {d u^\gamma} {ds}} +  \Gamma_{\alpha \beta}^\gamma
u^\alpha u^\beta \right)=
mc^2 {\frac {D u^\gamma} {ds}}  = 0,
\la{PEA}
\een
where ${\frac D {ds}}$ stands for the absolute derivative with respect
to the affine connection and
$u^\alpha = {\frac {d x^\alpha} {ds}}$.
In contrast the standard variational principle for the action
${\cal A}_m=- mc\int ds$ yields motion along a geodesic (G) lines
$mc^2\left( {\frac {d u^\gamma} {ds}} + { \gamma \brace \alpha \beta}
u^\alpha u^\beta \right) = mc^2 {\frac D {ds}} u^\gamma - {\cal F}^\gamma=
m\,a^\gamma - {\cal F}^\gamma=0$. Here a specific "torsion force"
${\cal F^\gamma} = 2 mc^2 {S^\gamma}_{\alpha \beta} u^\alpha u^\beta$
must be introduced to compensate the natural torsion dependence
of dynamics and to allow the free test particle to follow the extreme of
this classical action.
An analogous G-A problem we find too in field dynamics in spaces with torsion
even for scalar fields.

It will be very hard to drop out of the physical theory one of the two
fundamental principles: Newton's principle of inertia,
or the action principle based on quantum mechanics
\cite{DF} which conflict in presence of torsion.
Usually one prefers to preserve only the action principle (see the review
articles \cite{HehlR} and the huge amount of references therein).
But the theories of this type are not successful in description of
the physical reality (for recent results see \cite{BFY}).
Therefore it seems that the best way to overcome the G-A problem is to
modify the standard MCP and to look for a proper new one.

At first this idea was realized by A. Saa \cite{Saa} in pure geometrical way
for all type of fields in some special spaces with gradient
torsion vector $S_\alpha := {\frac 2 3} S_{\alpha\mu}{}^\mu$,\,
$S_\alpha = \partial_\alpha \Theta \equiv \Theta_{,\alpha}$.
We call the field $\Theta$ {\em a torsion dilaton}.
Saa suggested to replace the standard volume element $d^4 x \sqrt{|g|}$
with a new one: $d^4 x \sqrt{|g|} e^{-3\Theta}$.
Unfortunately this simple approach which yields A-type equations of motion
for fields turns to be unsuccessful for particles and fluids \cite{F1}
and contradicts to basic experimental facts of gravitational physics \cite{BFY}.
We can preserve the good features of this model in suitable way and look
for some more sophisticated self-consistent minimal coupling principle (SCMCP).
We may hope that SCMCP based on right physical reasons
will bring us after all to a consistent and beautiful theory
compatible with experimental facts.

A solution of the G-A problem for test particles,
analogous to Saa's model, was recently
proposed in \cite{KP}. It turns out that the simple action
\ben
{\cal A}_m=- mc\int e^{- \Theta} ds
\la{Am}
\een
yields precisely the A-type equations (\ref{PEA}) in the special case when
the torsion has the form
$S_{\alpha\beta}{}^\gamma = A_{\alpha\beta}{}^\gamma +
\Theta_{,[\alpha} \delta_{\beta]}^\gamma$
similar to the one used in string theory \cite{String},
$A_{\alpha\beta\gamma}:= S_{[\alpha\beta\gamma]}$ being the complete
anti-symmetric part of the torsion. In general
$S_{\alpha\beta}{}^\gamma = \Sigma_{\alpha\beta}{}^\gamma
+ A_{\alpha\beta}{}^\gamma + S_{[\alpha} \delta_{\beta]}^\gamma$,\,\,
$\Sigma_{\alpha\mu}{}^\mu\equiv 0$,\,\,$\Sigma_{[\alpha\beta\gamma]} \equiv 0.$
In this article we consider only the case when
$\Sigma_{\alpha\beta}{}^\gamma\equiv 0 $ and $S_\alpha \equiv \Theta_{,\alpha}$.
For brevity we call such torsion "a torsion of special basic form".

The corresponding action for relativistic fluid is \cite{F2}:
\ben
{\cal A}_\mu =
-\int d^4x \sqrt{|g|}\, e^{-\Theta}\left({\mu c^2 + \mu \Pi}\right)
\la{Amu}
\een
and yields the autoparallel equations of motion
\ben
(\varepsilon + p) u^\beta \nabla_\beta u_\alpha =
\left(\delta^\beta_\alpha - u_\alpha u^\beta \right)\nabla_\beta p.
\la{FluEA}
\een
Here $\Pi$ is the fluid's elastic energy, $\varepsilon$ is the fluid's
energy density, $p$ is the pressure, and $\mu= \mu_0
{\frac {\sqrt{g_{\mu \nu} {\dot x}^\mu {\dot x}^\nu}}  {J(x)\sqrt{|g|}}}$
is the fluid's mass density in Lagrange variables
(see for details \cite{F1} and the references there in).

To have an A-type equation of motion for scalar field $\varphi$ we need
to include a factor $e^{-3\Theta}$ in front of the corresponding kinetic term
${\sfrac 1 2}\,g^{\mu \nu} \partial_\mu \varphi \partial_\nu \varphi$
in the action.
To have a consistent with the particle action (\ref{Am}) semiclassical limit
of the equation for scalar field an additional
factor $e^{-2\Theta}$ in front of the mass term ${\sfrac 1 2}\,m^2 \varphi^2$\,
is needed \cite{F1}, \cite{F2}.
Hence, the action
\ben
{\cal A}_\varphi =
\int d^4x \sqrt{|g|}\,e^{-3\Theta}\,{\sfrac 1 2}\!
\left( g^{\mu \nu} \partial_\mu \varphi \partial_\nu \varphi -
m_\varphi^2 e^{-2\Theta}\varphi^2 \right)
\la{phiA0}
\een
will yield the A-type equation of motion:
\ben
\Box \varphi + m_\varphi^2 e^{-2\Theta} \varphi = 0
\la{NsfEA0}
\een
with semiclassical limit (i.e. $\varphi= A \exp({\frac i \hbar}S)$,\,\,
$m_\varphi ={\frac {m c} \hbar}$ and $\hbar \rightarrow 0$)
precisely the Hamilton-Jacobi equation for particle with
action (\ref{Am}):\,\,$g^{\alpha\beta}\,\partial_\alpha S\,\partial_\beta S
= m^2 e^{-2\Theta}$.

\section{Torsion Producing Dilatations and Torsion Dilaton}
In the spaces
${\cal M}^{4}\{g_{\alpha\beta}(x), \Gamma_{\alpha\beta}{}^\gamma(x)\}$\,\,
the action of the Weyl's local dilatations:
\ben
g_{\alpha\beta} \,\,\,{\stackrel{{}^{\sigma}}{\longrightarrow}}\,\,\,
g'_{\alpha\beta} = e^{2\sigma}g_{\alpha\beta}, \,\,\,\,\,
x^\alpha \,\,\,{\stackrel{{}^{\sigma}}{\longrightarrow}}\,\,\,
x'^\alpha \equiv x^\alpha = inv(\sigma)
\la{dil}
\een
with arbitrary function $\sigma(x)$ may be extended on the affine connection
$\Gamma_{\alpha\beta}{}^\gamma(x)$ (which owns independent degrees of freedom)
in different ways  \cite{Weyl},\cite{DTO}.
These non-coordinate transformations leave the local differentiable structure
of the manifold ${\cal M}^{(4)}$ unchanged and determine some mappings
\ben
{\cal M}^{4}\{g_{\alpha\beta}(x), \Gamma_{\alpha\beta}{}^\gamma(x)\}
\,\,\,{\stackrel{{}^{\sigma}}{\longrightarrow}}\,\,\,
{\cal M}^{4}\{g'_{\alpha\beta}(x), \Gamma'_{\alpha\beta}{}^\gamma(x)\}
\la{ManMap}
\een
changing the additional geometrical structures $g_{\alpha\beta}$
and $\Gamma_{\alpha\beta}{}^\gamma$.
We have some infinite dimensional group $G^\infty_\sigma$ which
preserves the angles but this property is not enough to define it.
There exist several similar groups $G^\infty_\sigma$ which
act on $\Gamma_{\alpha\beta}{}^\gamma$ in different ways.
Suitable for our purposes is the following extension of the action of the group
$G^\infty_\sigma$ on the affine connection \cite{DTO}:
\ben
\Gamma_{\alpha\beta}{}^\gamma
\,\,\,{\stackrel{{}^{\sigma}}{\longrightarrow}}\,\,\,
\Gamma'_{\alpha\beta}{}^\gamma =
\Gamma_{\alpha\beta}{}^\gamma+\sigma_{,\alpha}\delta_\beta^\gamma\,,\nonumber\\
S_{\alpha\beta}{}^\gamma
\,\,\,{\stackrel{{}^{\sigma}}{\longrightarrow}}\,\,\,
S'_{\alpha\beta}{}^\gamma =
S_{\alpha\beta}{}^\gamma  + \sigma_{,[\alpha}\delta_{\beta]}^\gamma
\la{dilG}
\een
which yields the important relations\footnote{The first of the relations
(\ref{dilRN}) was pointed out as a consequence of the first of the relations
(\ref{dilG}) already in \cite{Einstein}.}
\ben
R_{\alpha\beta\mu}{}^\nu
\,\,\,{\stackrel{{}^{\sigma}}{\longrightarrow}}\,\,\,
R'_{\alpha\beta\mu}{}^\nu\equiv
R_{\alpha\beta\mu}{}^\nu = inv_{{}_{TPD}}(\sigma), \nonumber \\
N_{\alpha\beta}{}^\gamma
\,\,\,{\stackrel{{}^{\sigma}}{\longrightarrow}}\,\,\,
N'_{\alpha\beta}{}^\gamma \equiv
N_{\alpha\beta}{}^\gamma = inv_{{}_{TPD}}(\sigma) \equiv0.
\la{dilRN}
\een
These show that such dilatations preserve the curvature and the
(zero) nonmetricity of the space
${\cal M}^{4}\{g_{\alpha\beta}(x), \Gamma_{\alpha\beta}{}^\gamma(x)\}$,
but produce an additional torsion $\sigma_{,[\alpha}\delta_{\beta]}^\gamma$.
Hence the name {\em torsion producing dilatations}
(TPD) and the notation $G_\sigma^\infty{}\!_{{}_{{}_{TPD}}}$.
It's easy to see that actually TPD change only the torsion vector:
\ben
\Sigma_{\alpha\beta}{}^\gamma = inv_{{}_{TPD}}(\sigma),\,\,\,\,\,
A_{\alpha\beta}{}^\gamma = inv_{{}_{TPD}}(\sigma),\nonumber \\
S_\alpha \,\,\,{\stackrel{{}^{\sigma}}{\longrightarrow}}\,\,\,
S'_\alpha = S_\alpha + \sigma_{,\alpha}.
\la{TorTPD}
\een

According to formulae (\ref{dil}), (\ref{dilG}) and (\ref{PEA}) the TPD
transformations change simultaneously the physical units (as usual
\cite{Weyl}), the parallel displacement and the free motion of the test
particles according to Newton law of inertia in the affine-metric space-time.

We need to make a few more steps in comparison with the articles \cite{DTO}.
The first is to extract the torsion dilaton $\Theta$ from torsion vector
$S_\alpha$ in a $G_\sigma^\infty{}\!_{{}_{{}_{TPD}}}$ invariant way
using the representation:
\ben
S_\alpha = \SD_\alpha + \Theta_{,\alpha},
\la{S_TD}
\een
${\ThetaD}_{,\alpha}$ being a solution of the gauge-fixing
condition $0 = g^{\mu\nu}{\SD}_\mu{\SD}_\nu=
g^{\mu\nu}(S_\mu- {\ThetaD}_{,\mu})(S_\nu- {\ThetaD}_{,\nu})$.
This partial differential equation is a kind of Hamilton - Jacobi equation
for a "massless charged particle" in an "electromagnetic field" with
vector-potential $S_\alpha$ and under curtain boundary conditions (for
example ${\ThetaD}(t_0,\vec{x})=0$) has an unique solution,
which fixes the vector $\SD_\alpha$ unambiguously and leads
to the following transformation rules:
\ben
{\SD}_\alpha = inv_{{}_{TPD}}(\sigma),\,\,\,\,\,
\Theta
\,\,\,{\stackrel{{}^{\sigma}}{\longrightarrow}}\,\,\,
\Theta' = \Theta + \sigma.
\la{S_TD_tr}
\een

Thus we see after all that the TPD really produce only torsion dilaton field
$\Theta$. Now we can introduce a new {\em dilaton connection} with coefficients
\ben
{\GammaD}_{\alpha\beta}{}^\gamma:=
\Gamma_{\alpha\beta}{}^\gamma - \Theta_{,\alpha}\delta_\beta^\gamma
= inv_{{}_{TPD}}(\sigma).
\la{GD}
\een

It's remarkable that it has the same curvature
${\RD}_{\alpha\beta\mu}{}^\nu= R_{\alpha\beta\mu}{}^\nu$
and the same parts of the torsion
${\SigmaD}_{\alpha\beta}{}^\gamma= \Sigma_{\alpha\beta}{}^\gamma$
and ${\AD}_{\alpha\beta}{}^\gamma= A_{\alpha\beta}{}^\gamma$
as the initial affine connection $\Gamma_{\alpha\beta}{}^\gamma$
and differs from it only with respect to the torsion vector
${\SD}_\alpha = S_\alpha - {\ThetaD}_{,\alpha}$. The torsion dilaton plays
the role of a non-standard compensating field with respect to the local
$G_\sigma^\infty{}\!_{{}_{{}_{TPD}}}$ mappings, as far as
${\GammaD}_{\alpha\beta}{}^\gamma = {\gamma \brace \alpha \beta}
-\delta_\beta^\gamma\Theta_{,\alpha}-\delta_\alpha^\gamma\Theta_{,\beta}
+ g_{\alpha\beta}\Theta^{,\gamma}+
\SD_{\alpha\beta}{}^\gamma + \SD{}^\gamma{}_{\alpha\beta} +
\SD{}^\gamma{}_{\beta\alpha}$.

Let $T_{\beta_1,\ldots\beta_n}{}^{\gamma_1,\ldots\gamma_m}$ be a (m,n)
tensor under the action of the infinite-dimensional group $G_x^\infty$ of
coordinate transformations with a TPD-dimension
$d_{{}_{T_{\beta_1,\ldots\beta_n}{}^{\gamma_1,\ldots\gamma_m}}}$,
i.e. it transforms under $G_\sigma^\infty{}\!_{{}_{{}_{TPD}}}$
mappings according to the rule
$T_{\beta_1,\ldots\beta_n}{}^{\gamma_1,\ldots\gamma_m}
\,\,\,{\stackrel{{}^{\sigma}}{\longrightarrow}}\,\,\,
T'_{\beta_1,\ldots\beta_n}{}^{\gamma_1,\ldots\gamma_m}=
e^{\sigma d_{{}_{T_{\beta_1,\ldots\beta_n}{}^{\gamma_1,\ldots\gamma_m}}}}
T_{\beta_1,\ldots\beta_n}{}^{\gamma_1,\ldots\gamma_m}$.
We define the action of a new {\em dilaton derivative}
$\nablaD_\alpha$ according to the formula:
\ben
\nablaD_\alpha T_{\beta_1,\ldots\beta_n}{}^{\gamma_1,\ldots\gamma_m} :=
\partial_\alpha T_{\beta_1,\ldots\beta_n}{}^{\gamma_1,\ldots\gamma_m} +
\nonumber \\
\GammaD_{\alpha\mu}{}^{\gamma_1}
T_{\beta_1,\ldots\beta_n}{}^{\mu,\ldots\gamma_m} + \ldots
- \GammaD_{\alpha\beta_1}{}^{\mu}
T_{\mu,\ldots\beta_n}{}^{\mu,\ldots\gamma_m}
\nonumber \\
 - \ldots  - d_{{}_{T_{\beta_1,\ldots\beta_n}{}^{\gamma_1,\ldots\gamma_m}}}
\Theta_{,\alpha}T_{\beta_1,\ldots\beta_n}{}^{\gamma_1,\ldots\gamma_m}=
\nonumber \\
\nabla_\alpha T_{\beta_1,\ldots\beta_n}{}^{\gamma_1,\ldots\gamma_m}-
\nonumber \\
\left(m-n+d_{{}_{T_{\beta_1,\ldots\beta_n}{}^{\gamma_1,\ldots\gamma_m}}}\right)
\Theta_{,\alpha}T_{\beta_1,\ldots\beta_n}{}^{\gamma_1,\ldots\gamma_m}.
\la{BarNabla}
\een

This definition together with formulae (\ref{S_TD_tr}), (\ref{GD})
and the standard transformation rule of any tensor under action
of the group $G_x^\infty$ show that the new derivative $\nablaD_\alpha$
is precisely the covariant derivative with respect to the product
$G_x^\infty \otimes G_\sigma^\infty{}\!_{{}_{{}_{TPD}}}$, i.e. the derivative
$\nablaD_\alpha T_{\beta_1,\ldots\beta_n}{}^{\gamma_1,\ldots\gamma_m}$ is
a (m,n+1) $G_x^\infty$ tensor with the TPD-dimension
$d_{{}_{\nabla^{\hskip -6.pt {}^{{}^{D}}}_{\alpha}
T_{\beta_1,\ldots\beta_n}{}^{\gamma_1,\ldots\gamma_m}}}
= d_{{}_{T_{\beta_1,\ldots\beta_n}{}^{\gamma_1,\ldots\gamma_m}}}$.
We call this new derivative a  co${}^2$variant derivative adopting
the terminology by Eddington and Dirac \cite{Weyl}.
 
As a basic consequences of the definition (\ref{BarNabla}) we derive
the following relations:
1) $\nablaD_\alpha T_{\beta_1,\ldots\beta_n}{}^{\gamma_1,\ldots\gamma_m}=
\nabla_\alpha T_{\beta_1,\ldots\beta_n}{}^{\gamma_1,\ldots\gamma_m}$ if
$d_{{}_{T_{\beta_1,\ldots\beta_n}{}^{\gamma_1,\ldots\gamma_m}}}= n - m$.
In particular $\ND_{\alpha\beta\gamma}:=\nablaD_\alpha g_{\beta\gamma}=
\nabla_\alpha g_{\beta\gamma}\equiv 0$ and this
permit us to commute the $\nablaD$ differentiation with the
lowering and raising of the indexes by metric $g_{\beta\gamma}$;\,\,
2) $\nablaD_\alpha\Theta \equiv 0$. Indeed, using the basic properties of the
linear connections \cite{SKN} and the relation $d_{e^{\Theta}}=1$ we obtain
$e^{\Theta}\nablaD_\alpha\Theta=\nablaD_\alpha \left(e^{\Theta}\right)=
\partial_\alpha \left(e^{\Theta} \right)-\Theta_{,\alpha} e^{\Theta}\equiv 0$.

\section{Self Consistent Minimal Coupling Principle in Presence
         of Torsion Dilaton}

\subsection{SCMCP for matter and a Novel TPD Symmetry}
For $u^\beta= {\frac {dx^\beta} {ds}}$ the values $d_{u^\beta}= -1, m=1, n=0$
imply $\nablaD_\alpha u^\beta \equiv \nabla_\alpha u^\beta$ and the A-type
equations (\ref{PEA}) and (\ref{FluEA}) may be rewritten
in a transparent co${}^2$variant way:
\ben
mc^2 {\frac {\DD u^\gamma} {ds}}  = 0\,, \nonumber \\
(\varepsilon + p) u^\beta \nablaD_\beta u_\alpha =
\left(\delta^\beta_\alpha - u_\alpha u^\beta \right)\nablaD_\beta p\,.
\la{NEW_{{}_E}M_P_F}
\een

These are A-type equations with respect to the dilaton connection $\nablaD$.
They fulfill the Newton's law of inertia and follow from action principle for
the actions (\ref{Am}) and (\ref{Amu}).
Hence, we have reached a SCMCP for matter.

The geometrical invariance of the solutions of the equations
(\ref{NEW_{{}_E}M_P_F}) under action of the group
$G_x^\infty \otimes G_\sigma^\infty{}\!_{{}_{{}_{TPD}}}$
is an obvious consequence of the corresponding invariance of the
actions (\ref{Am}) and (\ref{Amu})\, (where $d_{m}=0$
and $d_{\mu} = -3$)
\footnote{Our construction corrects the wrong statement about non-invariance of
the particle action under $G_\sigma^\infty$ mappings in the article \cite{KP}
where the TPD invariance of the A-lines was recognized for first time,
but a right understanding of the role of torsion dilaton
$\Theta(x)$ and of TPD parameters $\sigma(x)$ in it was not reached.}.
Moreover, it leads us to the important conclusion that our SCMCP yields
a {\em novel symmetry}: a TPD invariance of the matter dynamics
in spaces with torsion of the special basic form.
Taking into account Dirac's note \cite{Weyl}:
"It appears as one of the fundamental principles of Nature that
the equations expressing basic laws should be invariant under
widest possible group of transformations"
we will extend this new TPD symmetry to all matter fields, too.

\subsection{SCMCP and TPD Invariant Action for Gauge Fields}
As a basic example for gauge field $A_\alpha$  with gauge group
$G^\infty_\chi$:\,\,$A_\alpha
\,\,\,{\stackrel{{}^{\chi}}{\longrightarrow}}\,\,\,
A'_\alpha = A_\alpha + \chi_{,\alpha}$ we consider the electromagnetic field.
The corresponding generalization for non-abelian Yang-Mills fields is
straightforward. It's well known that Maxwell's equations
\ben
\partial_{[\mu} F_{\nu\lambda]} = 0, \,\,\,\,\, \partial_\mu D^{\mu\nu}= j^\nu
\la{Maxwell}
\een
are based only on the differentiable structure of the physical space-time
${\cal M}^{4}\{g_{\alpha\beta}(x), \Gamma_{\alpha\beta}{}^\gamma(x)\}$\,\,
and do not depend on its additional geometric structures
$g_{\alpha\beta}$ and $\Gamma_{\alpha\beta}{}^\gamma$.
A very clear explanation of this fundamental statement both from physical and
from mathematical point of view may be found in the recent article \cite{PLH}
together with complete list of references on this subject.
This fact becomes transparent when one writes down the equations (\ref{Maxwell})
using the Cartan's calculus of differential forms and exterior derivatives.
Hence, there is no need to generalize the equations (\ref{Maxwell}) in spaces
${\cal M}^{4}\{g_{\alpha\beta}(x), \Gamma_{\alpha\beta}{}^\gamma(x)\}$\,\,
using some covariant derivatives (which in presence of torsion would break
the gauge invariance). Then for gauge fields we have the
simplest possible MCP: their equation of motion in
${\cal M}^{4}\{g_{\alpha\beta}(x), \Gamma_{\alpha\beta}{}^\gamma(x)\}$\,\,
don't change at all being stable against any changes of its geometry \cite{PLH}.
For them is proper the geometry of the gauge space and the gauge fields play
the role of coefficients of another linear connection. If the base space
${\cal M}^{4}\{g_{\alpha\beta}(x), \Gamma_{\alpha\beta}{}^\gamma(x)\}$\,\,
of the gauge field theory has an affine-metric geometry
and the physics is  covariant with respect to the group
$G_x^\infty \otimes G_\sigma^\infty{}\!_{{}_{{}_{TPD}}}$,
we are to introduce a co${}^3$variant connection and derivative with
respect of the whole group $G_x^\infty \otimes
G_\sigma^\infty{}\!_{{}_{{}_{TPD}}}\otimes G^\infty_\chi$.

But in Maxwell's theory exists an important relation which depends on the
space-time geometry. This is the constitutive law
$D^{\mu\nu}= {\frac 1 2} \chi^{\mu\nu\rho\lambda} F_{\rho\lambda}$ with
some coefficients $\chi^{\mu\nu\rho\lambda}$.
If we chose the electromagnetic units so, that the electromagnetic factor is
${\frac {\varepsilon_0} {\mu_0}}=1$, in vacuum we have
$\chi^{\mu\nu\rho\lambda}= 2 \sqrt{|g|}g^{\rho[\mu}g^{\nu]\lambda}$ \cite{PLH}
and the Maxwell's equations (\ref{Maxwell}) follow from action principle with
familiar action
\ben
{\cal A}_A = - {\sfrac 1 4}\int d^4x D^{\mu\nu}F_{\mu\nu}=\nonumber \\
- {\sfrac 1 4}\int d^4x \sqrt{|g|} g^{\mu\rho}g^{\nu\lambda}
F_{\mu\nu}F_{\rho\lambda},
\la{A_A}
\een
$F_{\mu\nu}= A_{\mu,\nu}- A_{\nu,\mu}$ being the electromagnetic tensor
which plays the role of curvature in gauge space geometry.

The standard interaction of the gauge field $A_\alpha$ with classical
particle with charge $e$ is described by the action-term:
\ben
{\cal A}_{eA} = e \int A_\mu dx^\mu.
\la{AmeA}
\een

This term together with the free particle action (\ref{Am})
give a canonical momenta $p_\mu := \partial_{\dot x^\mu} L_{m,eA}=
p_0{}_\mu+ e A_\mu$,\,\,$ p_0{}_\mu:=
- {\frac {mc} {\sqrt{g_{\alpha\beta}\dot x^\alpha \dot x^\beta}}} e^{-\Theta}$
being the momenta of the free particle,
$\dot x^\mu:= {\frac {dx^\mu}{dt}}$.
Obviously $d_{\dot x^\mu}=0$ imply $d_{p_0{}_\mu} = 0$ and to have a
self-consistent theory we must impose the condition $d_{(e A_\mu)}= 0$.
The fundamental charges are discrete quantities and cannot change continuously,
hence $d_e=0$ which yields the conditions $d_{A_\mu}= 0$ and $d_{F_{\mu\nu}}=0$
and proves the TPD invariance of the actions (\ref{A_A}) and (\ref{AmeA}).
As a result we have SCMCP and TPD invariant dynamics for gauge fields.
Moreover, the dependence of the gauge field action (\ref{A_A}) on the geometry
of the space-time is of the same kind as the dependence of the matter actions
(\ref{Am}) and (\ref{Amu}), because in the action (\ref{A_A})
enter only exterior derivatives.

Some new generalizations of the constitute law of gauge fields theory
compatible with our SCMCP and with TPD invariance are possible in presence
of torsion. For example we can add to the action $G_x^\infty \otimes
G_\sigma^\infty{}\!_{{}_{{}_{TPD}}}\otimes G^\infty_\chi$ invariant terms
of the form
$\sqrt{|g|}e^{2\Theta}A^{\mu\nu\kappa}A_\kappa{}^{\rho\lambda}
F_{\mu\nu}F_{\rho\lambda}$, or
$\sqrt{|g|}e^{2\Theta}A^{\mu\nu\kappa}A_\kappa{}^{\rho\lambda}
F_{\mu\rho}F_{\nu\lambda}$.
We may extract from torsion pseudovector
${}^*\!A^\alpha:= \varepsilon^{\alpha\mu\nu\lambda}A_{\mu\nu\lambda}$
a pseudoscalar "axion" field $a(x)$ in the same 
$G_\sigma^\infty{}\!_{{}_{{}_{TPD}}}$ invariant way which gives the
torsion dilaton $\Theta$. Then we can construct a modified constitutive
law as described in  \cite{PLH}.
There exits too a $G_x^\infty \otimes
G_\sigma^\infty{}\!_{{}_{{}_{TPD}}}\otimes G^\infty_\chi$
invariant terms of other simple form, for example
$\sqrt{|g|} F_{\mu\nu}A^{\mu\nu\kappa}\SD_\kappa$.
The physical meaning of such interactions of gauge fields with space-time
metric-affine geometry is unclear and we give them only to
illustrate the fact that our SCMCP and the $G_x^\infty \otimes
G_\sigma^\infty{}\!_{{}_{{}_{TPD}}}\otimes G^\infty_\chi$ invariance are not
enough to fix these interactions.

\subsection{TPD Invariant Action and SCMCP for Massive Scalar Field}
The straightforward generalization of the action (\ref{phiA0}):
${\cal A}_\varphi^{\it D}= \int d^4x \sqrt{|g|}e^{-3\Theta}
{\sfrac 1 2}\left(g^{\mu \nu} \nablaD_\mu \varphi \nablaD_\nu \varphi-
m_\varphi^2 e^{-2\Theta} \varphi^2 \right)$ is
TPD invariant if $d_{\varphi}={\frac 1 2}$.\footnote{The TPD invariance
of this action implies $d_{m_\varphi}=0$.
In physical units $m_\phi={\frac {mc} \hbar}$.
Obviously (\ref{dil}) gives $d_c=0$. Then $d_m=0$ leads to $d_\hbar=0$ which
is in agreement with the known experimental data and will be used further.}
We may introduce a new scalar field $\phi = e^{-{\frac 3 2}\Theta}\varphi$
with normal dimension $d_{\phi}=-1$ and rewrite this action in a form:
\ben
{\cal A}_\phi =
\int d^4x \sqrt{|g|}{\sfrac 1 2}\!
\left( g^{\mu \nu} \nablaD_\mu \phi \nablaD_\nu \phi -
m_\phi^2 e^{-2\Theta}\phi^2 \right).
\la{Aphi}
\een
The corresponding TPD co${}^2$variant equation reads:
\ben
\BoxD \phi + m_\phi^2 e^{-2\Theta} \phi = 0.
\la{NsfEA}
\een
It's an A-type equation, semiclassically consistent with the action (\ref{Am}).
Hence, we have a SCMCP for the field $\phi$.

The requirement of TPD invariance introduces a non-minimal
interaction of the field $\phi$ with torsion dilaton since
$g^{\mu \nu} \nablaD_\mu \phi \nablaD_\nu \phi =
g^{\mu \nu}(\phi_{,\mu} + \Theta_{,\mu}\phi)
(\phi_{,\nu} +\Theta_{,\nu}\phi)$,
but there is no non-minimal coupling with metric $g_{\mu\nu}$.
This property will be of crucial importance when one will try to construct
a TPD invariant theory with scalar fields $\phi$ preserving the general
relativity (GR) as a right theory of the metric part of space-time geometry.
In contrast, to reach a conformal invariant equation for scalar field
in Riemannian spaces one must introduce a non-minimal coupling
with metric \cite{PCT}.
In presence of torsion dilaton we have an additional TPD
invariant term: ${\frac 1 6}\sqrt{|g|} R_\Theta\phi^2=
inv_{{}_{TPD}}(\sigma)$ with
$ R_\Theta = {\stackrel{{}^{\{\}}}{R}} +
6 \left( {\stackrel{{}^{\{\}}}{\Box}} \Theta -
g^{\mu\nu} \Theta_{,\mu} \Theta_{,\nu} \right) $\,being the Cartan scalar
curvature when $\SD_{\alpha\beta}{}^\gamma \equiv 0$,\,
${\stackrel{{}^{\{\}}}{\Box}}=
g^{\alpha\beta} \stackrel{{\{\}}}{\nabla}_\alpha \stackrel{{\{\}}}{\nabla}_\beta
={1\over \sqrt{|g|}}\partial_\mu\left(\sqrt{|g|}g^{\mu\nu}\partial_\nu\right)=
\Box + 3 S^\mu \nabla_\mu$ being the Laplace-Beltrami operator,
${\stackrel{{}^{\{\}}}{R}}$ being the corresponding Riemann curvature.
The Penrose-Chernicov-Tagirov action with this term added with arbitrary
constant coefficient $\rho$ gives the most general form of a quadratic in
the field $\phi$ TPD invariant action:
\ben
{\cal A}_{\phi;\rho} =
\int d^4x \sqrt{|g|}{\sfrac 1 2}\!
\biggl( g^{\mu \nu}  \phi_{,\mu} \phi_{,\nu} \nonumber + \\
{\sfrac 1 6} \Bigl({\stackrel{{}^{\{\}}}{R}} + \rho R_\Theta\Bigr)\phi^2 -
m_\phi^2 e^{-2\Theta}\phi^2\biggr).
\la{Aphi2}
\een
In it ${\frac 1 6} \Bigl({\stackrel{{}^{\{\}}}{R}}+\rho R_\Theta\Bigr)=
{\frac 1 6}(1+\rho){\stackrel{{}^{\{\}}}{R}}+
\rho\Bigl( {\stackrel{{}^{\{\}}}{\Box}} \Theta -
g^{\mu\nu} \Theta_{,\mu} \Theta_{,\nu} \Bigr)$ and
the unique choice $\rho=-1$ yields the only TPD invariant action
$ {\cal A}_{\phi;-1} =
\int d^4x \sqrt{|g|}{\sfrac 1 2}\!
\Bigl( g^{\mu \nu}  \phi_{,\mu} \phi_{,\nu} -
\Bigl({\stackrel{{}^{\{\}}}{\Box}}\Theta-g^{\mu\nu}\Theta_{,\mu} \Theta_{,\nu}
+ m_\phi^2 e^{-2\Theta}\Bigr)\phi^2 \Bigr)$
with {\em minimal coupling} of the scalar field $\phi$ with metric
$g_{\alpha\beta}$. It coincides with the action (\ref{Aphi})
up to divergence term and gives immediately the following
transparent form of the equation (\ref{NsfEA}):\,
$ {\stackrel{{}^{\{\}}}{\Box}} \phi +
\Bigl({\stackrel{{}^{\{\}}}{\Box}}\Theta- g^{\mu\nu}\Theta_{,\mu} \Theta_{,\nu}
+ m_\phi^2 e^{-2\Theta}\Bigr)\phi = 0$.

\subsection{TPD Invariant Action and SCMCP for Dirac Spinor Field}
To incorporate the Dirac spinor field $\psi$ in our consideration we need
coefficients of the spin-connection ${\GammaD}_{\alpha b}{}^c$.
We introduce the usual tetrads $e^\alpha_a$, $e^a_\alpha$
and extend the action of the derivative
$\nablaD_\alpha$ on them using the spin-connection $\GammaD_{\alpha b}{}^c$:
$\nablaD_\alpha e^\beta_b = \partial_\alpha e^\beta_b
+ \GammaD_{\alpha\mu}{}^{\beta} e^\mu_b -
\GammaD_{\alpha b}{}^c e^\beta_c - d_{e^\beta_b}\Theta_{,\alpha} e^\beta_b$, \,
$\nablaD_\alpha e_\beta^b = \partial_\alpha e_\beta^b
- \GammaD_{\alpha\beta}{}^{\mu} e_\mu^b +
\GammaD_{\alpha c}{}^b e_\beta^c - d_{e_\beta^b}\Theta_{,\alpha} e_\beta^b$.
Tetrads are fixed by the condition
$g_{\alpha \beta}= e^a_\alpha e^b_\beta \eta_{ab}$\,($\eta_{ab}$ being the
Minkowski's metric) up to
(pseudo)orthogonal rotation. We justify them imposing usual additional
conditions $d_{e^a_\alpha}=-d_{e_a^\alpha}=1$\, and
$\nablaD_\alpha e^\beta_b \equiv 0$,\, $\nablaD_\alpha e_\beta^b \equiv 0$\,
which define the spin-connection coefficients in form:
\ben
\GammaD_{\alpha b}{}^c = \GammaD_{\alpha\beta}{}^{\gamma} e^\beta_b e_\gamma^c
- e^\beta_b \partial_\alpha e_\beta^c + \Theta_{,\alpha} \delta^c_b.
\la{sp_con}
\een
This formula and relations  (\ref{S_TD_tr}), (\ref{GD}) show that
$\GammaD_{\alpha b}{}^c \equiv \Gamma_{\alpha b}{}^c=inv_{{}_{TPD}}(\sigma)$,\,
$\Gamma_{\alpha b}{}^c$ being the spin-connection coefficients of the initial
affine connection.

Let us consider Dirac field $\psi$ using standard $\gamma$-matrices. We have
$d_{\gamma^a}=d_{\gamma_a}=0$, $d_{\gamma^\alpha}= -d_{\gamma^\alpha}=1$
and $d_\psi=-{\frac 3 2}$. Now it's easy to check that the TPD invariant action
\ben
{\cal A}_\psi =
\int d^4x \sqrt{|g|}\biggl({\sfrac i 2}\!
\Bigl( \bar\psi \gamma^\alpha (\nablaD_\alpha\psi ) -
(\nablaD_\alpha\bar\psi ) \gamma^\alpha \psi \Bigr) - \nonumber \\
m_\psi e^{-\Theta}\bar\psi\psi\biggr)
\la{Apsi}
\een
produces the A-type Dirac equation
\ben
i \gamma^\alpha \nablaD_\alpha\psi - m_\psi e^{-\Theta} \psi = 0
\la{DiracE}
\een
with right semiclassical limit.
Here $\nablaD_\alpha\psi = \partial_\alpha\psi -
{\sfrac 1 4}\GammaD_{a b c}\sigma^{bc}\psi +  {\sfrac 3 2}\Theta_{,\alpha}\psi$
is the spinor $G_x^\infty \otimes G_\sigma^\infty{}\!_{{}_{{}_{TPD}}}$
co${}^2$variant derivative ($\sigma^{\alpha\beta}=
\gamma^{[ \alpha} \gamma^{\beta ]}$). In presence of gauge field
$A_\alpha$ the corresponding
$G_x^\infty \otimes G_\sigma^\infty{}\!_{{}_{{}_{TPD}}}\otimes G_\phi^\infty$
co${}^3$variant derivative for a charged spinor field $\psi$ will include
the standard additional term $-i e A_\alpha\psi$.
Hence, we have SCMCP for spinor field.

A similar considerations convince us that our SCMCP may be extended on matter
fields with all values of the spin. Hence, having universal character it
gives a proper solution of the G-A problem for all fields
if torsion has a special basic form.

\section{Action for Geometrical Fields and Violation of
the TPD Symmetry}
We can choose the action for geometrical fields in a form, similar to the
one for gauge fields (\ref{A_A}). At present days level of experimental accuracy
the GR is known to be a right theory of the metric part of space-time
geometry \cite{Will}. Hence, to comply with the experimental data we have to
preserve the metric dependence of the Hilbert-Einstein action in presence of
torsion. This correspondence principle gives "a gravitation constitutive law"
in the form $D^{\mu\nu\alpha\beta}=
-{\frac {2c} {\kappa}} \sqrt{|g|}g^{\alpha[\mu}g ^{\nu]\beta} F_G(\Theta)$
($\kappa$ being Einstein constant),
as far as the torsion dilaton field $\Theta$ is the {\em only scalar field}
which enters in the total torsion tensor ${S_{\alpha\beta} }^\gamma$
being independent of the metric $g_{\alpha\beta}$. Hence:
\ben
{\cal A}_G =-{\sfrac 1 4}\int d^4x D^{\mu\nu\alpha\beta}R_{\mu\nu\alpha\beta}=
\nonumber \\
-{\sfrac c {2\kappa} } \int d^4 x \sqrt{|g|}\,F_G(\Theta)\,R
\la{AG}
\een
with an arbitrary positive new function $F_G(\Theta)$.
This way we worked out the action for the torsion dilaton $\Theta$
(which enters in the Cartan curvature $R$), too
without putting by hands additional terms and {\em dimensional} coupling
constants. Obviously we have a specific form of SCMCP for the very torsion
dilaton field and the space-time geometry will be complete determined
by the usual properties of the matter without any new "charges".
The dimensionless multiplier $F_G(\Theta)$ in the action (\ref{AG}) shows that
the torsion dilaton $\Theta$ may be considered as a kind of Brans-Dicke field
$\Phi(x) = F_G(\Theta(x))$ and when $\SD_{\alpha\beta}{}^\gamma \equiv 0$
our model falls in the well known class of the scalar-tensor theories
of gravity \cite{BD}, but with a new specific geometric role of
the scalar field $\Theta$.

If we are going to take into account the nonzero cosmological constant
$\lambda_c$, a term
\ben
{\cal A}_c = -{\sfrac c {2\kappa} }\int d^4x \sqrt{|g|}\,2\lambda_c F_c(\Theta)
\la{Ac}
\een
with one more arbitrary dimensionless function $F_c(\Theta)$ must be added
to the action (\ref{AG}).

Obviously formulae (\ref{AG}) and (\ref{Ac}) in general bring us to a model
with variable gravitational and cosmological constants which are governed by
torsion dilaton field. In the space-time domains where $\Theta \approx const$
this model resembles GR with rescaled by corresponding factors
$F_G^{-1}(\Theta)$ and $F_c(\Theta)$ gravitational constant $\kappa$ and
cosmological constant $\lambda_c$.

Unfortunately the SCMCP gives no instructions about the choice of the functions
$F_G(\Theta)$ and $F_c(\Theta)$. We have to justify them in some new way.
The simple choice $F_G(\Theta)= e^{-2\Theta}$, $F_c(\Theta)= e^{-4\Theta}$
yields a TPD invariance of the actions (\ref{AG}) and (\ref{Ac}).
Then in Einstein frame
($ g^{\mu \nu} = e^{-2\Theta} g^{\mu \nu}_{E}$ -- see Appendix A) we will
turn back to the usual GR due to the TPD invariance of the total action
${\cal A}_{tot} = {\cal A}_G+{\cal A}_c+{\cal A}_M$, ${\cal A}_M = {\cal A}_m
+ {\cal A}_\mu+{\cal A}_A+{\cal A}_{eA}+{\cal A}_\phi+{\cal A}_\psi+\ldots$
being the action of the whole matter which is TPD invariant by construction.
Hence, in Einstein frame this variant of theory is consistent with all known
gravitational experimental data \cite{Will}, with exception of these
connected with early-Universe which have not satisfactory description in
the original GR, and in it the torsion dilaton disappears as
{\em a physical field}.

More interesting is to use a functions $F_G(\Theta)\neq e^{-2\Theta}$
and (or) $F_c(\Theta)\neq e^{-4\Theta}$
which {\em violates in a soft way} the TPD invariance of the
whole theory\footnote{It was shown that action similar to (\ref{AG})
yields cosmological models without singularities in pure dilatonic gravity
for a large class of functions $F_G(\Theta)$ \cite{Rama}
(see \cite{CRM}, too).}.
Now, going to the Einstein frame
($g^{\mu \nu}= F_G(\Theta) g^{\mu \nu}_{E}$ -- see Appendix A)
we have to replace the torsion
dilaton field with a new {\em physical dilaton field} $D(x)$
defined as $D = \Theta + {\sfrac 1 2}\ln F_G(\Theta)$ at all places where
the torsion dilaton  $\Theta$ was standing in the exact TPD invariant action
${\cal A}_M= \int d^4 x\, \LambdaD_M$.
The action of geometrical fields with cosmological constant
term which may be rewritten as
\ben
{\cal A}_{G,c}= \int d^4 x \Lambda_{G,c} =
\int d^4 x \left( e^{2D}\,\LambdaD_G + e^{4D}\,V\,\LambdaD_c\right)
\la{A_Gc}
\een
($\LambdaD_G := -{\sfrac c {2\kappa}} \sqrt{|g|}\,e^{-2\Theta}\, R $ and
$\LambdaD_c := -{\sfrac c \kappa} \sqrt{|g|}\,e^{-4\Theta}\,\lambda_c$,
being the corresponding TPD invariant Lagrange densities, $V := F_c/F_G^2$
being a dimensionless "cosmological potential" for the field $\Theta $)
in Einstein frame takes the form
\ben
{\cal A}_{G,c}^{\!{}^E} = -{\sfrac c {2\kappa}}
\int d^4 x \sqrt{|g_{{}_E}|} \left(R_{{}_E} + 2 \lambda_c V_{{}_E}(D)\right)=
\nonumber \\
-{\sfrac c {2\kappa}} \int d^4 x \sqrt{|g_{{}_E}|}
\Bigl({\stackrel{{}^{\{\}}}{R}}_{{}_E}
- 6 g^{\mu\nu}_{{}_E} D_{,\mu} D_{,\nu} \nonumber \\
- A^{\!{}^E}{}_{\alpha\beta\gamma} A_{{}_E}^{\alpha\beta\gamma} +
2 \lambda_c V_{{}_E}(D) \Bigr)
+ \oint_{\partial {\cal M}^{4}}\!\!...
\la{AGcE}
\een
where "$\dots$" denote a term essential for the TPD properties of the
action ${\cal A}_G$ but not for the complete set of field equations:
\ben
{\stackrel{{}^{\{\}}}{R}}_{{}_E}{}_{\alpha\beta} -
{\sfrac 1 2}{\stackrel{{}^{\{\}}}{R}}_{{}_E} g^{\!{}^E}_{\alpha\beta}
- 6 \left( D_{,\alpha} D_{,\beta} -
{\sfrac 1 2}g^{\!{}^E}_{\alpha\beta}g^{\mu\nu}_{{}_E} D_{,\mu} D_{,\nu}\right)
\nonumber \\
-  \lambda_c  V_{{}_E}(D) g^{\!{}^E}_{\alpha\beta} =
{\sfrac \kappa {c^2}} T^{\!{}^E}_{\alpha\beta},
\nonumber \\
{\stackrel{{}^{\{\}}}{\Box}}{}^{\!{}^E} D + {\sfrac 1 6}\lambda_c V'_{{}_E}(D) =
{\sfrac \kappa {6 c^2} } T^{\!{}^E},
\nonumber \\
{\frac {\delta\LambdaD_M{}^{\hskip -9pt{}^E}}{\delta x^\alpha_{{}_{fluid}}}}=0,
\,\,\,
{\frac {\delta\LambdaD_M{}^{\hskip -9pt{}^E}} {\delta A_\alpha}}=0,
\,\,\,
{\frac {\delta\LambdaD_M{}^{\hskip -9pt{}^E}} {\delta \phi}}=0,
\,\,\,
{\frac {\delta\LambdaD_M{}^{\hskip -9pt{}^E}} {\delta \psi}}=0,
\,\,\, \dots
\la{FEq}
\een
where the following quantities are taken in Einstein frame:
$T_{\alpha\beta} := { \sfrac {2c} {\sqrt{|g|}} }
{\frac{\delta\LambdaD_M}{\delta g^{\alpha\beta}}}$
-- the energy-momentum tensor of the matter,
$T = g^{\alpha\beta}T_{\alpha\beta}$ being its trace,
$V'_{{}_E}(D)= {\sfrac {dV_{{}_E}} {dD}}$ -- the derivative
of the cosmological potential $V_{{}_E}(D)= V(\Theta(D))$ and
in addition the condition $A_{\alpha\beta\gamma}=0$ is imposed.

As we see, the source of the physical dilaton field is the trace
of the energy-momentum tensor. Hence, if we wish to allow a nonzero trace
of energy-momentum tensor we must introduce a nonzero physical dilaton
$D \neq 0$, i.e. we must use a function $F_G \neq e^{-2\Theta}$.

The next important conclusion is that the physical dilaton $D$
for which we have nontrivial dynamical equation,
absorbs the arbitrary function $F_G$.
As a result in the model under consideration actually
presents only one arbitrary function -- the cosmological potential $V(D)$
which has to be determined using independent physical reasons.
If we suppose $V(D) \equiv 1$, we will have in all Einstein-frame-relations
the usual cosmological-constant-$\lambda_c$-terms. But there exist other
possibilities, which may turn to be more interesting from physical point
of view. For example, one may try to use the cosmological potential
of proper type and the physical dilaton field for description of the inflation
mechanism in the Universe (see \cite{Inflation} and the references therein).
This possibility is a new reason to consider space-time geometries more
general then Riemannian one and to enlarge the geometrical framework of GR.
On the other hand the geometrical meaning of the scalar field $\Theta$
may help us in fixing its potential.

In Dicke frame ($g^{\mu \nu} = e^{-2\Theta} g^{\mu \nu}_{\!{}_{\cal D}}$,
$D\neq 0$ -- see Appendix A) the action for geometrical fields has a form
\ben
{\cal A}_{G,c}^{\cal D}=-{\sfrac c {2\kappa}}\int d^4 x\sqrt{|g_{{}_{\cal D}}|}
\bigl(\Phi_{{}_{\cal D}} R_{{}_{\cal D}} +
2 \lambda_c V_{{}_{\cal D}}(\Phi_{{}_{\cal D}})
\bigr),
\la{A_GcD}
\een
where $\Phi_{{}_{\cal D}}(\Theta) = e^{2\Theta}F_G(\Theta)=e^{2D}$
is a Brans-Dicke field which replaces the torsion dilaton $\Theta$ and
$V_{{}_{\cal D}}(\Phi_{{}_{\cal D}}) :=
e^{4\Theta(\Phi_{{}_{\cal D}})} F_c(\Theta(\Phi_{{}_{\cal D}}))$.
Hence, in ${\cal D}$-frame our model reduces to standard Brans-Dicke model
with $\omega \equiv 0$ \cite{BD}, but in addition we have an arbitrary
cosmological potential $V_{{}_{\cal D}}(\Phi_{{}_{\cal D}})$.
If $\SD_{\alpha\beta}{}^\gamma \equiv 0$ the matter action ${\cal A}_M$
and matter equations of motion will have their GR form
and $R_{{}_{\cal D}} \equiv {\stackrel{{}^{\{\}}}{R}}{{}_{\cal D}}$
even if $D \neq 0$. In this case we arrive to a scalar-tensor theory of
gravity with scalar potential \cite{BD}, \cite{Inflation}
and geometrical fields' equations acquire the known form:
\ben
\Phi_{{}_{\cal D}}\!\left(
{\stackrel{{}^{\{\}}}{R}}_{{}_{\cal D}}{}_{\alpha\beta} -
{\sfrac 1 2}{\stackrel{{}^{\{\}}}{R}}_{{}_{\cal D}}
g^{\!{{}_{\cal D}}}_{\alpha\beta}\right)
- {\stackrel{{}^{\{\}}}{\nabla}}_\alpha{}^{\hskip -7pt {}_{\cal D}}
  {\stackrel{{}^{\{\}}}{\nabla}}_\beta {}^{\hskip -7pt {}_{\cal D}}
  \Phi_{{}_{\cal D}}
+ g^{\!{{}_{\cal D}}}_{\alpha\beta}
{\stackrel{{}^{\{\}}}{\Box}}{}^{\!{{}_{\cal D}}} \Phi_{{}_{\cal D}}
\nonumber \\
-  \lambda_c  V_{{}_{\cal D}}(\Phi_{{}_{\cal D}})
    g^{\!{}_{\cal D}}_{\alpha\beta} =
{\sfrac \kappa {c^2}} T^{{{}_{\cal D}}}_{\alpha\beta},
\nonumber \\
{\stackrel{{}^{\{\}}}{\Box}}{}^{\!{{}_{\cal D}}}\Phi_{{}_{\cal D}} +
{\sfrac 2 3}\lambda_c \left(\Phi_{{}_{\cal D}}
   V'_{{}_{\cal D}}(\Phi_{{}_{\cal D}})
- 2 V_{{}_{\cal D}}(\Phi_{{}_{\cal D}})\right) =
{\sfrac \kappa {3 c^2} } T^{{{}_{\cal D}}}.
\la{FEq_D}
\een
Now using the well known properties of the scalar-tensor theories of gravity
\cite{Hanlon} we see that under standard week requirements on the cosmological
potential our model will be consistent with the gravitational experiments in
solar system, as far as that without cosmological term (\ref{Ac})
it is equivalent to original Brans-Dicke theory with $\omega = 0$
and contradicts to these experiments. We see, too that the physical dilaton
may be considered precisely as a scalar field one needs for inflation scenario,
(see \cite{Inflation} and the references therein).

All consistent with the observations theories of gravity are not conformal
invariant. The present model of gravity with propagating torsion gives in
addition a new basis for introducing scalar-tensor theories of gravity using
a fine tuned violation the TPD symmetry. For example, a Brans-Dice kinetic
term with $\omega \neq 0$ for the torsion dilaton may be incorporated in our
consideration, too but it would violate more strongly the TPD symmetry
because the transformation law (\ref{S_TD_tr}) and the identity
$\nablaD_\alpha\Theta \equiv 0$ do not allow a TPD invariant kinetic terms
for the field $\Theta$, different from the one we have already in the action
(\ref{AG}). Our approach shows that the very Brans-Dicke model may be considered
as a model with TPD symmetry, violated into two steeps: first one is the soft
violation we use in present article, it depends only on the values of the
field $\Theta$, and the second one -- the introduction of the Brans-Dicke
kinetic term with $\omega \neq 0$ which depends on the gradients of this field.
Hence, the values $\omega \neq 0$ for Brans-Dicke parameter may be accepted
only under pressure of some new experimental facts.

\section{Cartan Scalar Curvature and Cosmological Constant}
The last result we wish to report in this article may turn to be quite
interesting in the light of the recent development of the cosmological
constant problem \cite{ICHEP}, as far as for determination of the cosmological
potential for dilaton field.
If $D\neq const$ the TPD invariance of all lagrangian
densities $\LambdaD$ and the field equations after some algebra  give
the important relation (see Appendix B):
\ben
R + 4\lambda_c \bigl({\sfrac 1 4} V'(\Theta) + V(\Theta)\bigr) F_G(\Theta) = 0.
\la{Rlr}
\een
This pure geometrical relation is complete independent of the matter
and its motion in contrast to the corresponding one in GR:
$\stackrel{{}^{\{\}}}{R} + 4 \lambda_c = {\sfrac \kappa {c^2}}T$ because the
trace of the energy-momentum tensor, being related with the physical
dilaton according to dynamical equations of our model,
enters in the Cartan scalar curvature.
Thus for first time we are able to recover the physical meaning of
the Cartan scalar curvature of the space-time relating it with
the cosmological constant and with cosmological term
(\ref{Ac}) in the total action.

The TPD (\ref{dil}) with $\bar\sigma= {\sfrac 1 2}\ln \left(F_G(\Theta)
\left|{\sfrac 1 4} V'(\Theta) + V(\Theta)\right|\right)$
brings us to a simplest form of the relation (\ref{Rlr}):
$\bar R = \mp 4 \lambda_c = const$ {\em in presence of matter of any kind},
$\pm \equiv sign\bigl({\sfrac 1 4} V'(\Theta)+V(\Theta)\bigr)$. We see that
in this "cosmological constant" representation of our model of gravity with
propagating torsion the Cartan scalar curvature, being constant,
has no singularities, independently of the presence of matter and its motion.
As far as the test particle trajectories in our model are A-lines (\ref{PEA}),
such property may confirm the conclusions of the early articles \cite{Esposito},
but a more detailed analysis of the behavior of the whole Cartan curvature
tensor and of the corresponding tidal forces is needed to clarify
the problem with space-time singularities.

It is interesting to discuss the form of the relation between Cartan scalar
curvature and cosmological constant in different frames (see Appendix A).
In the Einstein frame the relation (\ref{Rlr}) takes the form
$R_{{}_E} + 4 \lambda_c({\sfrac 1 4} V'_{{}_E}(D) + V_{{}_E}(D)) = 0$
and may be derived directly from field equations (\ref{FEq}).
In Dicke frame we have the same relation in a form
$R_{{}_{\cal D}} + 2 \lambda_c V'_{{}_{\cal D}}(\Phi_{{}_{\cal D}}) = 0$.
It may be obtained directly by variation of the action (\ref{A_GcD})
with respect to the Brans-Dicke field $\Phi_{{}_{\cal D}}$.
Now we see that the Cartan scalar curvature $R$ would be zero in any frame
if $\lambda_c = 0$. When $\lambda_c \neq 0$ is arbitrary, Cartan scalar
curvature $R$ may be zero only in some specific frame if the cosmological
potential has corresponding specific form:
i) $V_0(\Theta) = e^{-4\Theta}$ -- in basic frame;
ii) $V_{0{}_E}(D)) = e^{-4D}$ -- in Einstein frame;
iii) $V_{0{}_{\cal D}}(\Phi_{{}_{\cal D}}) = 1 $ -- in Dicke frame.
When $\lambda_c \neq 0$ we will have more general relations:
i) $R = -4\lambda_c$ in basic frame if
$ V(\Theta)= 1 + v\,e^{-4\Theta}\int d\Theta e^{4\Theta}F^{-1}_G(\Theta)$;
ii) $R_{{}_E} = -4 \lambda_c$ if $V_{{}_E}(D))= 1 + v\, e^{-4D}$
-- in Einstein frame;
iii)$R_{{}_{\cal D}} = -4 \lambda_c$ if
$V_{{}_{\cal D}}(\Phi_{{}_{\cal D}})= 2 \Phi_{{}_{\cal D}} + const$.
In all these cases we have a variant of the
present model of gravity with torsion in which no arbitrary functions present
and the only parameter to be determined from the experimental data
is the cosmological constant $\lambda_c$, and maybe some additional constant
$v$ in the cosmological potential. Some of these cosmological potentials
have been investigated with respect of the inflation scenario \cite{Inflation}.
These examples show that our geometrical approach may offer new
reasons for fixing this potential which is the main open physical problem.

The properties discussed in this section as far as the very existence
of the physical dilaton $D$ are direct consequences of the soft violation of
TPD symmetry in our model.

\section{Concluding remarks}
We have to mention the following additional features and problems
of the present model of gravity with propagating torsion:

1) Due to the TPD invariance of the matter action
in this model we have a modified {\em equivalence principle}, as far as
at each point $x_0$ of the space-time exist a preferable local frame
in which $g_{\alpha\beta}(x_0)= \eta_{\alpha\beta}$,\,
$\GammaD_{\alpha\beta}{}^\gamma(x_0)=0$ \cite{Hartley}
and dynamical equations for matter, {\em being all of autoparallel type},
will simultaneously take their special relativistic form up to terms
of higher order as in GR. This may be reached in Dicke frame
by proper choice of the coordinates as in GR. If we wish to have such
generalized equivalence principle, we must preserve the TPD invariance of
the matter action ${\cal A}_M$ violating the TPD symmetry.

2) The further development of the present model may be based on its obvious
similarity to string models (see equation (\ref{AGcE})).
If it will be possible to relate our dilaton fields $\Theta$ and $D$
with string dilaton, we may justify the form of the action in our model.
For example the term
$A^{\!{}^E}{}_{\alpha\beta\gamma} A_{{}_E}^{\alpha\beta\gamma}$ in action
(\ref{AGcE}) may be related with the string axion
(see \cite{Hammond} and the references therein).

On the other hand our approach gives a new possibility to relate
the string dilaton with space-time torsion, not with nonmetricity, as one 
usually does in string theory \cite{String}.
As we sow, our consideration gives definite predictions for dilaton
interactions with usual matter.
At present this is an open problem in string theory.
Most probably our A-type equations of motion for matter in presence of torsion
dilaton give the framework for the results of a supersymmetry soft violation
in superstring theories, expected to produce the mass spectrum of the
real matter in them.

3) A detailed analysis of the experimental consequences of the present model
is needed, as far as of the possibility to incorporate in it
a torsion of more general form then the special basic one,
or to use more complicated violation schemes of the new TPD symmetry.
More deep understanding of geometrical and physical meaning of this
symmetry and of the TPD transformations is desirable, too.

\bigskip
\bigskip
\bigskip

\noindent{\Large\bf Acknowledgments}
\bigskip

The author wishes to express his thanks for the stimulating conversations
on different parts of this article to V.~de~Alfaro, R.~A.~Asanov,
B.~M.~Barbashov, M.~Cavaglia, N.~A.~Chernicov, A.~T.~Filippov, D.~I.~Kazakov,
V.~V.~Nesterenko, A. Pelster, E.~A.~Tagirov, A.~A.~Zheltukhin.
He is indebted to S.~Yazadjiev for many useful discussions on the subject
under consideration and especially on the relation of the present model of
gravity with Brans-Dicke theory.
The author is grateful to Y.~N.~Obuchov for pointing out the hardly available
book: V.~N.~Ponamarev, A.~O.~Borovinskii, Y.~N.~Obuchov,
{\em Geometro-dinamicheskie metody i kalibrovochny podhod k teorii
gravitacionnyh vzaimodeistvii}, Energoatomizdat, Moscow, 1985 (in Russian)
where one can find detailed consideration of Einstein-Cartan theory.
This work has been partially supported by
the Sofia University Foundation for Scientific Researches, Contract~No.~245/98,
and by
the Bulgarian National Foundation for Scientific Researches, Contract~F610/98.
The author is grateful, too to the leadership of the Bogoliubov Laboratory
of Theoretical Physics, JINR, Dubna, Russia for hospitality and
working conditions during his stay there in the summer of 1998
when the main part of this investigation has been performed.

\section{Appendix}

\subsection{Weyl's Changes of the Frame}

As usual (see \cite{Weyl}, \cite{DTO}, \cite{CRM} and the references therein),
in the present article we use the Weyl's conformal transformations (\ref{dil})
without change of the torsion $S_{\alpha\beta}{}^\gamma$ and nonmetricity
$N_{\alpha\beta}{}^\gamma$ as transformations which change the frame.
This way we introduce another type of {\em curvature producing dilatations}
(CPD). They look more familiar and belong to a group of local gauge changes
$G_\sigma^\infty{}\!_{{}_{{}_{CPD}}}$
with transformation rule\, $\Gamma_{\alpha\beta}{}^\gamma
\,\,\,{\stackrel{{}^{\sigma}}{\longrightarrow}}\,\,\,
\Gamma'_{\alpha\beta}{}^\gamma = \Gamma_{\alpha\beta}{}^\gamma +
\sigma_{,\alpha}\delta_\beta^\gamma + \sigma_{,\beta}\delta_\alpha^\gamma-
g_{\alpha\beta}\sigma^{,\gamma}$\,
which resembles the transformation law of the Christoffel symbols.
For them we have
$R_{\alpha\beta\mu}{}^\nu \longrightarrow R'_{\alpha\beta\mu}{}^\nu \neq
R_{\alpha\beta\mu}{}^\nu \neq inv_{{}_{CPD}}(\sigma)$,\,
$S_{\alpha\beta}{}^\gamma=S'_{\alpha\beta}{}^\gamma=inv_{{}_{CPD}}(\sigma)$,\,
$N_{\alpha\beta}{}^\gamma = N'_{\alpha\beta}{}^\gamma=inv_{{}_{CPD}}(\sigma)$.

Then for the TPD invariant quantities $\LambdaD(g_{\alpha\beta},\Theta;...)$
the identity $\LambdaD (g_{\alpha\beta},\Theta;...) \equiv
\LambdaD (e^{2\sigma} g_{\alpha\beta}, \Theta + \sigma;...)$
gives the CPD-transformation rule
$\LambdaD (g_{\alpha\beta},\Theta;...)=
\LambdaD (e^{-2\sigma} g'_{\alpha\beta}, \Theta;...)=
\LambdaD (g'_{\alpha\beta}, \Theta + \sigma;...)$.

In general we consider three different frames:

i) the basic one with metric $g_{\alpha\beta}$, torsion dilaton $\Theta$,...
(no additional frame-indexes are used)
where the initial considerations has been performed.

ii) the Einstein frame (E-frame) where all quantities carry the index "E".
By definition $\Lambda_G (g_{\alpha\beta},\Theta;...):=
-{\sfrac c {2\kappa}} \sqrt{|g|}\,F_G(\Theta)\,R(g_{\alpha\beta},\Theta;...)=:
e^{2D} \LambdaD_G (g_{\alpha\beta},\Theta;...) =
-{\sfrac c {2\kappa}} \sqrt{|g_{{}_E}|} \left({\stackrel{{}^{\{\}}}{R}}_{{}_E}
+ \ldots \right)$.
Hence, the transition from basic frame to E-frame
corresponds to the value $\sigma_{\!{}_E} = D-\Theta$
and all TPD invariant quantities in E-frame
reach the form $\LambdaD (g_{\alpha\beta},\Theta;...)=
\LambdaD (g^E_{\alpha\beta}, D;...)$.

iii) the Dicke frame (${\cal D}$-frame) where all quantities carry the index
"${\cal D}$". The transition from basic frame to ${\cal D}$-frame
corresponds to the value $\sigma_{\!{}_{\cal D}} = -\Theta$.
Hence in ${\cal D}$-frame all TPD invariant quantities reach the form
$\LambdaD (g_{\alpha\beta},\Theta;...)=
\LambdaD (g^{\!{}_{\cal D}}_{\alpha\beta}, 0;...)$.
If the TPD invariant part of the torsion equals zero,
these quantities will have their GR form even in the case $D \neq 0$
(otherwise these quantities have the same form as in Einstein-Cartan theory
of gravity \cite{HehlR}). This happens precisely in the case of matter action
${\cal A}_M$ and matter equations of motion in our model.
In the case $D \equiv 0$ the ${\cal D}$-frame coincides with the E-frame.

\subsection{Derivation of the Relation between Cartan Scalar Curvature and
Cosmological Constant}

Here we work in the basic frame where the total Lagrange density is
$$\Lambda_{tot} = \Lambda_G+\Lambda_c+\Lambda_M =
e^{2D}\LambdaD_G + e^{4D}V(D)\LambdaD_c + \LambdaD_M.$$
From the field equations:
${\sfrac {\delta \Lambda_{tot}} {\delta g_{\mu\nu}}}=0$
and
${\sfrac {\delta \Lambda_{tot}} {\delta \Theta}}=0$
using the identities: a)
$2 g_{\mu\nu}{\frac{\delta\LambdaD}{\delta g_{\mu\nu}}} +
{\frac{\delta\LambdaD}{\delta \Theta}} \equiv 0$ which describes in
infinitesimal form the TPD-invariance of the quantities $\LambdaD$;
and b) ${\sfrac {\delta D} {\delta g_{\mu\nu}}}\equiv 0$ which describes
the independence of the torsion dilaton from metric,
it is not hard to obtain ${\sfrac {dD} {d\Theta} }
\left(2\Lambda_G + (4+{\sfrac {V'} V}) \Lambda_c \right) = 0$.
If $D\neq const$ this dynamical identity, fulfilled on the solutions of the
field equations, yields the relation $R + \lambda_c (V'+ 4 V)F_G = 0$
which reflects the TPD-properties of the model under consideration.

\end{document}